\documentclass[12pt,psfig,amsfonts]{article}

\input{epsf}

\begin{document}

\begin{center}
{\bf \Large Chiral Mesons in Hot Matter}
\\
\vspace*{0.5cm}
 A.~G\'omez Nicola, F.J.~LLanes-Estrada and J.R.Pel\'aez\\
Departamentos de F\'{\i}sica Te\'orica I y II, \\
Facultad de Ciencias F\'{\i}sicas,  Universidad Complutense, 28040
Madrid, Spain \end{center}


 \begin{abstract} We review our recent work on thermal meson
properties within the Chiral Perturbation Theory framework. We
will focus on the pion electromagnetic form factor, stressing its
importance for Relativistic Heavy Ion Collisions. We obtain
model-independent predictions (based only on chiral symmetry) such
as the temperature dependence of the pion electromagnetic charge
radius. Imposing  unitarity allows to describe the thermal effects
of resonances such as the $\rho$ meson.
\end{abstract}

\vspace*{0.5cm}

\section{Introduction}
In order to describe the meson gas formed after a Relativistic
Heavy Ion Collision (RHIC) for temperatures below  Chiral Symmetry
Restoration, Chiral Perturbation Theory (ChPT) is the most general
theoretical scheme \cite{gale87} and its predictions are based
only on Chiral Symmetry. When applied to the low-$T$ meson gas,
one can analyze static quantities such as the pressure
\cite{gele89,dope9902}, quark condensate
\cite{gale87,gele89,dope9902} and pion decay constants
\cite{gale87,fpifourth}. All these quantities point towards Chiral
Symmetry Restoration. In addition, the analysis of the pion
dispersion relation $p^2=m_\pi^2+g(p_0,\vert \vec{p} \vert;T)$
shows that to leading order $O(T^2)$, $g$ is real and depends only
on temperature but not on momenta \cite{gale87} while corrections
 to the pion width start to $O(T^4)$ \cite{fpifourth,gole89}.

 Recently, we have shown that  ChPT is equally useful to
 describe dynamical quantities such as scattering amplitudes
 \cite{glp02} or form factors \cite{glp04}. Furthermore, combining
  Chiral Symmetry Breaking  and
 Unitarity provides a fruitful  approach to describe  the thermal
behaviour
 of resonances \cite{dglp02} such  as the $\rho$, whose medium
properties are crucial to explain  the
 excess observed in the dilepton spectrum data \cite{dileptonexp},
 which has motivated a huge amount of theoretical work devoted to
 explain in-medium resonances  \cite{rhoteor}.

\section{Pion form factors at finite temperature}
The main contribution to the dilepton rate comes from the pion
annihilation channel via a virtual photon exchange
$\pi^+\pi^-\rightarrow \gamma^*\rightarrow e^+ e^-$. The dilepton
rate is then proportional to the pion electromagnetic form factor
\cite{kaj86gaka88} which for $T\neq 0$ has the following general
form:

\begin{eqnarray}
    \langle \pi^+(p)\pi^-(p')\vert V_0(0)\vert 0\rangle &=&
    q_0 F_t(S_0,\vert \vec{S} \vert,q_0;T)\nonumber\\
\langle \pi^+(p)\pi^-(p')\vert V_k(0)\vert 0\rangle &=&
    q_k F_s(S_0,\vert \vec{S} \vert,q_0;T)+S_k
    q_0 G_s(S_0,\vert \vec{S} \vert,q_0;T)
    \label{ffgenform}
    \end{eqnarray}
with $V_\mu$ the electromagnetic current,  $S=p+p'$, $q=p-p'$ and
$F_t$, $F_s$, $G_s$  even functions in $q_0$ by charge conjugation
invariance. These functions are related by the gauge invariance
Ward identity  $\langle \pi\pi\vert \partial_\mu
V^\mu\vert0\rangle=0$.

\begin{figure}[ht]
\centerline{\epsfxsize=2.8in\epsfbox{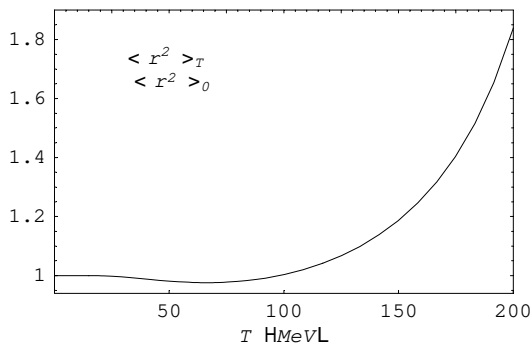}}
\caption{The pion electromagnetic charge radius at finite
temperature \label{fig:rad}}
\end{figure}

We have calculated the one-loop ChPT   form factors, whose
detailed results can be found in
 \cite{glp04}. This allows for  model-independent predictions.
 For instance, defining the pion electromagnetic charge and radius as:
\begin{eqnarray}
  Q_T&=&\lim_{\vert \vec{S} \vert\rightarrow 0^+}F_t(0,\vert \vec{S}
\vert;T)\nonumber\\
 \langle r^2 \rangle_T&=&-\frac{6}{Q_T}\left.\frac{d
F_t(0,\vert \vec{S} \vert;T)}{d \vert \vec{S}
\vert^2}\right\vert_{\vert \vec{S} \vert=0}
\end{eqnarray}
which takes into account Lorentz covariance breaking at $T\neq 0$,
one finds the result shown in Figure \ref{fig:rad}. The radius
remains almost constant for low $T$, which is consistent with the
Vector Meson Dominance (VMD) idea that $\langle r^2 \rangle \sim
M_\rho^{-2}$ \cite{res}, provided $M_\rho$ changes little at low
temperatures. On the other hand, the radius increases
significantly for higher temperatures, dominated by a $Q_T$
reduction (charge screening). Estimating the critical temperature
of deconfinement by requiring that the electromagnetic pion volume
$V(T)=(4\pi/3)\langle r^2\rangle_T^{3/2}$ equals the inverse pion
density gives $T_c\simeq 200$ MeV, very close to the prediction of
ChPT for chiral symmetry restoration
\cite{gale87,gele89,dope9902}. This gives support to the idea that
both transitions take place nearly at the same point. Our results
for the pion charge radius confirm those in \cite{dom} based on
QCD sum rules.

\section{Thermal unitarity and resonances}
In the center of mass frame ($\vec{p}=-\vec{p'}$, corresponding to
back to back dilepton pairs) the ChPT pion form factor satisfies
the following perturbative unitarity relation  for energies above
the two-pion threshold $S_0>2m_\pi$:

\begin{equation}
 \mbox{Im} F^{(1)}(S_0+i\epsilon;T)=\sigma_T(S_0) a_{11}^{(0)} (S_0^2)
\left[F^{(0)}(S_0^2)\right]^*\label{pertunit}\end{equation} with:
\begin{equation}
\sigma_T (S_0)=\sqrt{1-\frac{4m_\pi^2}{S_0^2}}
\left[1+\frac{2}{e^{S_0/2T}-1}\right]
\end{equation}

The superscripts $(0)$ and $(1)$ indicate the LO and NLO ChPT
contributions respectively and $a_{IJ}$ denote pion scattering
partial waves with definite isospin $I$ and angular momentum $J$ .
Note that $F^{(0)}=1$ but we have displayed it in the above
equation for consistency with what follows. Here, $\sigma_T$ is
the thermal two-particle phase space, which can be understood by
writing it as $\sigma_T=\sigma_0
\left[(1+n_B)(1+n_B)-n_B^2\right]$ where
$n_B=[\exp(S_0/2T)-1]^{-1}$ is the Bose-Einstein distribution
function for each pion. Hence, the phase space is enhanced at
finite temperature due to the difference between induced emission
and absorption processes in the thermal bath \cite{weldon}. This
effect contributes to the resonance thermal width enhancement
 \cite{dglp02}. The partial waves $a_{IJ}$ also satisfy a thermal
 perturbative unitarity relation \cite{glp02} which is the $T=0$ one
 with $\sigma_0$ replaced by
  $\sigma_T$, as in (\ref{pertunit}).

\begin{figure}[ht]
\centerline{\epsfxsize=3in\epsfbox{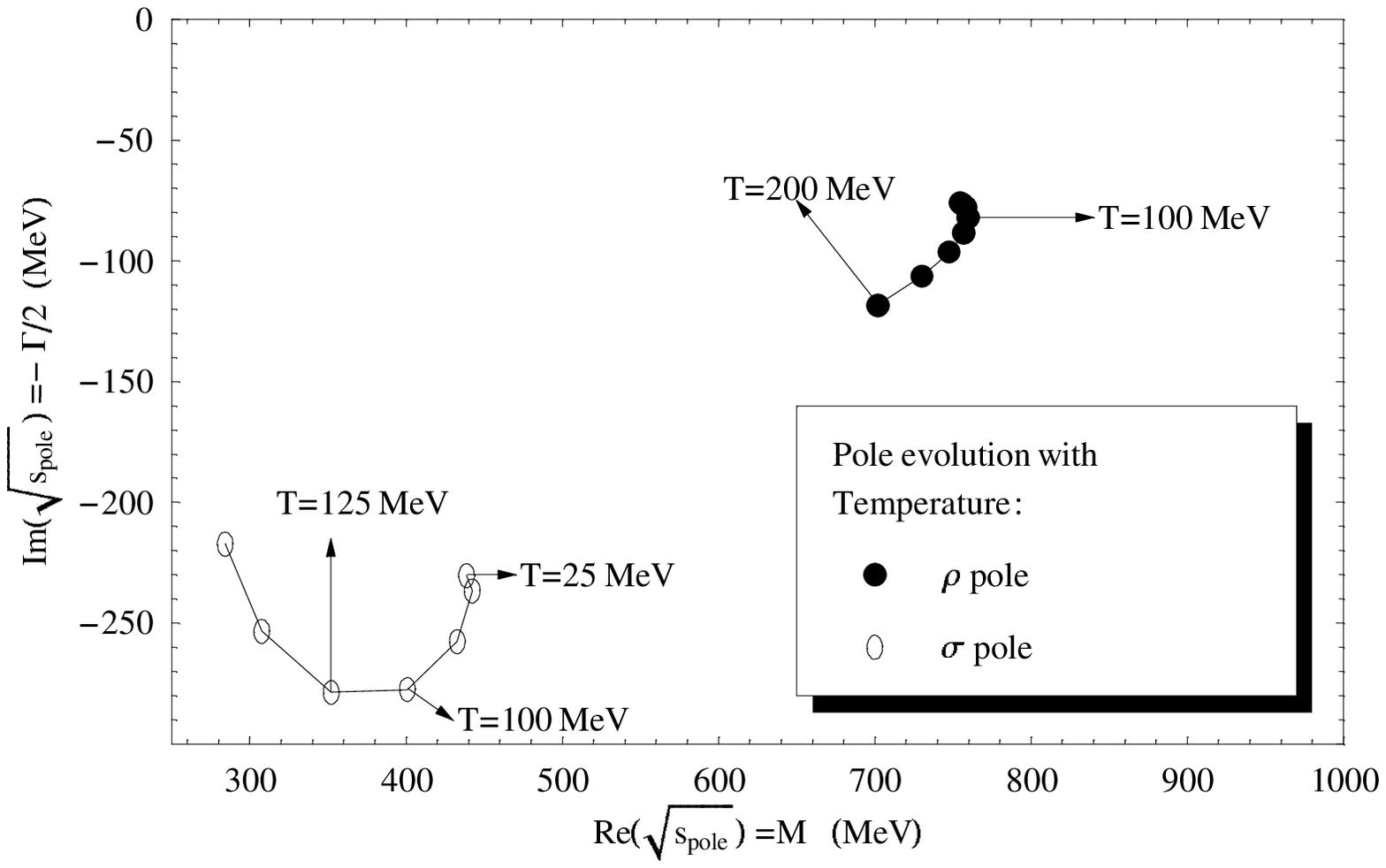}\epsfxsize=2.9in
\epsfbox{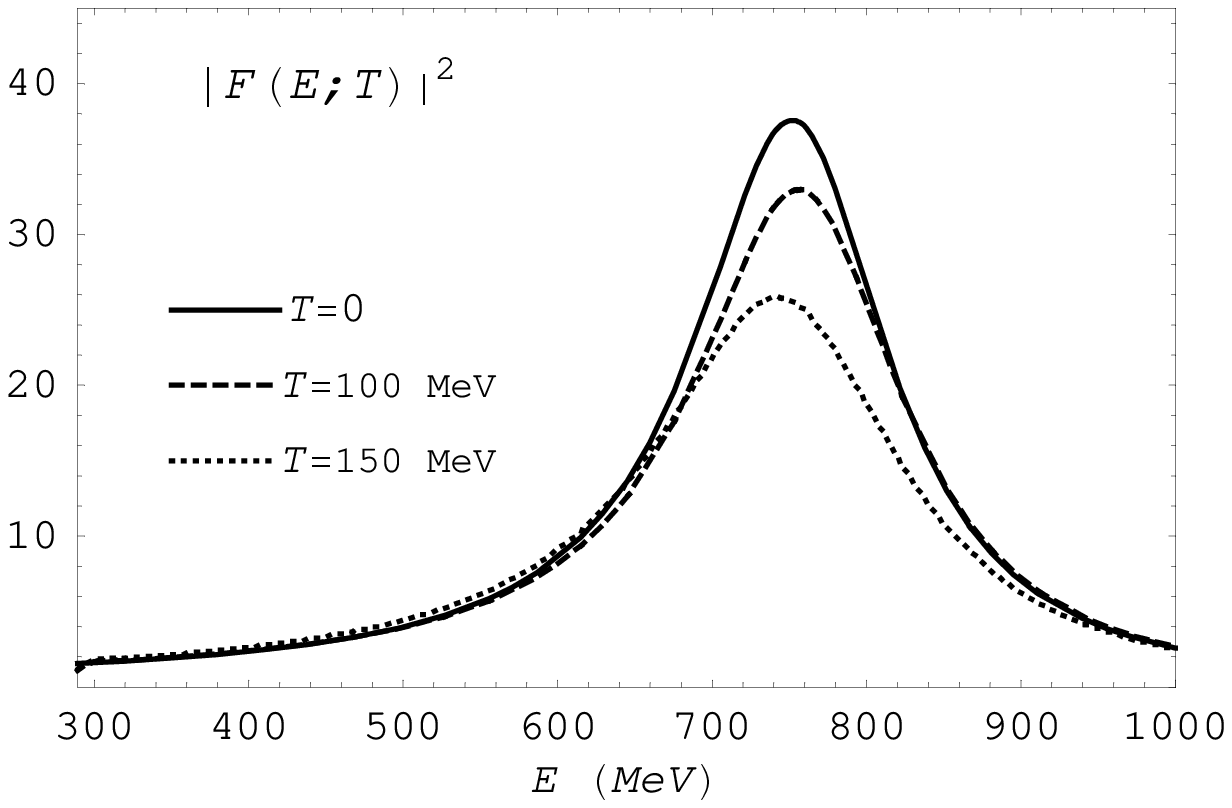}} \caption{Thermal poles  (left
pannel) and unitarized form factor (right pannel).
\label{fig:poles}}
\end{figure}

Thermal perturbative unitarity is the key point to describe
resonances within ChPT. In fact, inspired by the success of the so
called Inverse Amplitude Method (IAM) at $T=0$ \cite{iam} we have
constructed unitarized partial waves and form factors demanding
{\em exact} thermal unitarity. For instance,  $\mbox{Im}
F^{IAM}=\sigma_T a_{11}^{IAM} \left[F^{IAM}\right]^*$. The
remaining freedom is fixed by matching the ChPT series at low
energies. This approximation is valid for small pion densities
$n_B$,  which is very reasonable for the temperatures and energies
we are interested in. The IAM generates the $\sigma$ ($I=J=0$) and
$\rho$ ($I=J=1$) poles in the energy complex plane and their
thermal behaviour is the expected one. In Figure \ref{fig:poles}
we have plotted the results for the poles and for the unitarized
form factor. The $\rho$ widens  in the thermal bath, effect also
visible in the form factor, peaked at the $\rho$ mass. Such
widening is in perfect agreement with the expectations from
dilepton experiments  and previous  analysis
\cite{rhoteor,soko96}. Our resulting $M_\rho$ remains almost
constant for $T<$ 100 MeV (consistently with VMD) and decreases
slightly for higher $T$. This result points in the same direction
as recent measurements by the STAR collaboration \cite{star} but
we have ignored other medium effects such as baryon density. The
in-medium $\rho$ mass remains certainly an open issue
\cite{rhoteor}. As for the $\sigma$ pole, the results in Figure
\ref{fig:poles} show a behaviour compatible with Chiral Symmetry
Restoration: $m_\sigma$ decreases sharply, hence approaching
$m_\pi$ which increases softly with $T$ \cite{gale87}. The
$\sigma$ width increases at first by phase space enhancement but
as $m_\sigma\rightarrow 2m_\pi$ the decay $\sigma\rightarrow 2\pi$
is reduced and so does the width.
 A similar behaviour has been found in \cite{hat}.

Summarizing, ChPT and its unitarization programme can be extended
successfully to finite temperature and provide a promising
framework for future applications in the context of Relativistic
Heavy Ion Collisions.

\section*{Acknowledgments}
This work is supported by  the Spanish research projects
FPA2000-0956,PB98-0782, BFM2000-1326 and FPA2004-02602.


\vspace*{0.5cm}

\begin{thebibliography}{99}

\bibitem{gale87}
J.~Gasser and H.~Leutwyler,
{\it Phys.Lett.}  {\bf B184}, 83 (1987).

\bibitem{gele89}
P.Gerber and H.Leutwyler, {\it Nucl.Phys.}{\bf B321}, 387 (1989).

\bibitem{dope9902} A.Dobado,  and J.R.Pel\'aez, {\it Phys.Rev} {\bf D59}, 034004
(1999); J.R.Pel\'aez, {\it Phys.Rev} {\bf D66}, 096007 (2002).


\bibitem{fpifourth}
R.D.Pisarski and M.Tytgat, {\it Phys.Rev.} {\bf D54}, 2989 (1996).
D.Toublan, {\it Phys.Rev.} {\bf D56}, 5629 (1997). J.M.Martinez
Resco and M.A.Valle Basagoiti, {\it Phys.Rev.} {\bf D58}, 097901
(1998).

\bibitem{gole89}
J.L. Goity and H.Leutwyler {\it Phys.Lett.} {\bf B228}, 517
(1989).

\bibitem{glp02}
A.~G\'omez Nicola, F.~J.~Llanes-Estrada and J.~R.~Pel\'aez, {\it
Phys.Lett.} {\bf B550}, 55 (2002).

\bibitem{glp04} A.~G\'omez Nicola, F.~J.~Llanes-Estrada and
J.~R.~Pel\'aez, hep-ph/0405273.

\bibitem{dglp02}
A.~Dobado, A.~G\'omez Nicola, F.~Llanes-Estrada and
J.~R.~Pel\'aez, {\it Phys.Rev.} {\bf C66}, 055201 (2002).



\bibitem{dileptonexp} G.Agakichiev {\it et al}
[CERES Collaboration]
 {\it Phys. Rev. Lett.} {\bf 75}, 1272 (1995); {\it Phys.Lett.}
{\bf B442}, 405 (1998); {\it Nucl.Phys.} {\bf A 661}, 23 (1999);
D.Adamova {\it et al} [CERES/NA45 Collaboration] , {\it Phys. Rev.
Lett.}, {\bf 91} 042301 (2003).




\bibitem{rhoteor} M.Dey, V.L.Eletsky and B.L.Ioffe, {\it Phys.Lett.} {\bf B252}, 620
(1990); R.D.Pisarski, {\it Phys.Rev.} {\bf D52}, 3773 (1995);
G.~Q.~Li, C.~M.~Ko and G.~E.~Brown, {\it Phys.Rev.Lett.}  {\bf
75}, 4007 (1995). V. Koch and C. Song, {\it Phys. Rev.}  {\bf
C54}, 1903 (1996).
 R.Rapp and J.Wambach, {\it Adv.Nucl.Phys} {\bf 25}:1
(2000); V.~L.~Eletsky, M.~Belkacem, P.~J.~Ellis and J.~I.~Kapusta,
{\it Phys. Rev.} {\bf C64}, 035202 (2001). H.-J. Schulze and
D.Blaschke, {\it Phys.Lett.} {\bf B386}, 429 (1996); {\it
Part.Nucl.Lett.} {\bf 119}, 27 (2004).




\bibitem{kaj86gaka88} K.Kajantie, J.Kapusta, L.McLerran and A.Mekjian, {\it Phys.Rev.} {\bf D34}, 2746
(1986). C.Gale and J.I.Kapusta, {\it Phys.Rev.} {\bf C35}, 2107
(1987)

\bibitem{soko96} C.Song and V.Koch, {\it Phys.Rev.} {\bf C54} 3218 (1996).

\bibitem{res}
J.F.Donoghue, C.Ramirez and G.Valencia, {\it Phys.Rev.} {\bf D39},
1947 (1989);
 G.~Ecker, J.~Gasser, A.~Pich and E.~de Rafael, {\it Nucl.Phys.} {\bf B321}, 311 (1989).


\bibitem{dom} C. A. Dominguez, M. Loewe and J. S. Rozowsky, {\it Phys.Lett.}
{\bf B335}, 506 (1994).

\bibitem{weldon} H. A. Weldon, {\it Ann. Phys.} {\bf 214}, 152 (1992).

\bibitem{iam} T. N. Truong, {\it Phys.Rev.Lett.}{\bf 61}, 2526  (1988);
 A. Dobado, M.J.Herrero and T.N. Truong, {\it Phys.Lett.} {\bf B235}, 134 (1990); A.
Dobado and J.R. Pel\'aez, {\it Phys.Rev.}{\bf D47}, 4883 (1993);
{\it Phys.Rev.}{\bf D56}, 3057 (1997); A.Dobado, M.J.Herrero,
J.R.Pel\'aez and E.Ruiz Morales, {\it Phys.Rev.} {\bf D62}, 055011
(2000); A.G\'omez Nicola and J.R.Pel\'aez, {\it Phys.Rev.}{\bf
D65}:054009 (2002).

\bibitem{star}
J.~Adams {\it et al.}  [STAR Collaboration], {\it Phys.Rev. Lett.}
{\bf 92}:092301 (2004).

\bibitem{hat} S.Chiku and T.Hatsuda, {\it Phys.Rev.}{\bf D57}, R6
(1998); {\it Phys.Rev.}{\bf D58}, 076001 (1998). K.Yokokawa,
T.Hatsuda, A.Hayashigaki  and T.Kunihiro, {\it Phys.Rev.} {\bf
C66}, 022201 (2002).

\end{thebibliography}
\end{document}